\documentclass[traditabstract]{aa}

\usepackage{graphicx}
\usepackage{txfonts}
\usepackage{longtable}
\usepackage{natbib}
\bibpunct{(}{)}{;}{a}{}{,}
\begin{document}

\title{Binary energy source of the HH~250 outflow and its circumstellar environment
\thanks{Based on observations obtained with the VLT (Cerro Paranal, Chile) in programs 089.C-0196(A), 095.C-0488(A), and 095.C-0488(B), as well as with IRTF (Mauna Kea, Hawaii), SMA (Mauna Kea ,Hawaii), and the Nordic Optical Telescope (La Palma, Canary Islands, Spain)}
}
\author{Fernando Comer\'on\inst{1}
\and Bo Reipurth\inst{2}
\and Hsi-Wei Yen\inst{3}
\and Michael S. Connelley\inst{2}{\thanks{Staff Astronomer at the Infrared Telescope Facility, which is operated by the University of Hawaii under contract NNH14CK55B with the National Aeronautics and Space Administration}}
}

 \institute{
  European Southern Observatory, Alonso de C\'ordova 3107, Vitacura, Santiago, Chile\\
  \email{fcomeron@eso.org}
  \and
  Institute for Astronomy, University of Hawaii at Manoa, 640 N. Aohoku Place, Hilo, HI 96720, USA
  \and
  European Southern Observatory, Karl-Schwarzschild-Strasse 2, 85748, Garching bei M\"unchen, Germany
  }
%
%\offprints{}
%\mail{}
%\titlerunning{}
%\authorrunning{}
%\date{\today}
%
\date{Received; accepted}
%
%\abstract
% context heading (optional) %
%{}
% aims heading (mandatory)
%{aims}
% methods heading (mandatory)
%{methods}
% results heading (mandatory)
%{results}
% conclusions (optional)
%{conclusions}

%Abstract begin
\abstract {
%{\em Context.}  \\
{\em Aims.} Herbig-Haro flows are signposts of recent major accretion and outflow episodes. We aim to determine the nature and properties of the little-known   outflow source HH~250-IRS, which is embedded in the Aquila clouds. \\
{\em Methods.} We have obtained adaptive optics-assisted $L$-band images with the NACO instrument on the Very Large Telescope (VLT), together with $N-$ and $Q-$band imaging with VISIR also on the VLT. Using the SINFONI instrument on the VLT we carried out $H-$ and $K-$band integral field spectroscopy of HH~250-IRS, complemented with spectra obtained with the SpeX instrument at the InfraRed Telescope Facility (IRTF) in the $JHKL$ bands. Finally, the SubMillimeter Array (SMA) interferometer was used to study the circumstellar environment of HH~250-IRS at 225 and 351~GHz with CO~(2-1) and CO~(3-2) maps and 0.9~mm and 1.3~mm continuum images.\\
{\em Results.} The HH~250-IRS source is resolved into a binary with 0$\farcs 53$ separation, corresponding to 120~AU at the adopted distance of 225~pc. The individual components show heavily veiled spectra with weak CO absorption indicative of late-type stars. Both are Class~I sources, but their spectral energy distributions between 1.5~$\mu$m and 19~$\mu$m differ markedly and suggest the existence of a large cavity around one of the components. The millimeter interferometric observations indicate that the gas mainly traces a circumbinary envelope or disk, while the dust emission is dominated by one of the circumstellar envelopes.\\
{\em Conclusions.} HH~250~IRS is a new addition to the handful of multiple systems where the individual stellar components, the circumstellar disks and a circumbinary disk can be studied in detail, and a rare case among those systems in which a Herbig-Haro flow is present.}
%Abstract end


\maketitle
%________________________________________________________________

\section{Introduction} \label{intro}

The majority of stars are members of binary systems \citep[e.g.][]{Duquennoy91, Raghavan10}. Over the past 25 years it has been documented that the binary fraction is enhanced among T~Tauri stars compared to the field \citep[e.g.][]{Reipurth93,Duchene13}, and increases for Class~I objects \citep{Connelley08a,Connelley08b} and even further for Class~0 objects \citep{Chen13}. \citet{Larson72} suggested that all stars may be born in small multiple systems, which decay through dynamical interactions into the mixture of single, binary, and higher-order systems observed among field stars. This idea finds significant support from both observations and numerical simulations \citep[e.g.][]{Reipurth14}. When a triple system decays, the usual outcome is that the lightest member is expelled into a distant orbit or into an escape, while the remaining newly bound binary shrinks. Since this process in at least half the cases occurs during the embedded phase, then the hardened binary evolves in a viscous environment leading to orbital decay \citep{Reipurth10}. The increasingly frequent periastron passages cause accretion events and thus outflow episodes. \citet{Reipurth00} suggested that the dynamical evolution of newly formed binaries can explain the structure of giant Herbig-Haro (HH) flows, and demonstrated that HH energy sources very often are binaries. We have undertaken a survey of outflow sources and high-accretion objects with the aim to find and study binaries in this population. In the present paper we present a detailed study of the driving source of the little-known HH~250 flow in Aquila, largely based on new, high angular resolution near-infrared observations that show it to be a resolved binary. We also present millimeter and submillimeter interferometric observations of the surroundings of the binary, showing the different distributions of gas and dust that we interpret in terms of a combination of circumstellar and circumbinary disks.

\section{Observations and data reduction} \label{obs}

\begin{table*}[t]
\caption{Summary of observations}
\begin{tabular}{llll}
\hline
Instrument & Band & Date & Mode \\
\hline
SINFONI (VLT) & $H$ and $K$ & 24/25 August 2015 & Integral field spectroscopy \\
NACO (VLT)    &     $L'$    & 18/19 July 2012 & Imaging \\
VISIR (VLT)   & $N$         & 20/21 July 2015 & Imaging \\
VISIR (VLT)   & $Q$         & 6/7 August - 3/4 September 2015 & Imaging \\
SpeX (IRTF)   & 0.7 to 4.2~$\mu$m & 17/18 June 2015 & Spectroscopy \\
SMA           & 0.9 mm & 31 March 2015 & Integral field spectroscopy\\
SMA           & 1.3 mm & 27 February and 6 March 2015 & Integral field spectroscopy\\
EMMI(NTT)     & H$\alpha$   & 18/19 August 1994 & Imaging \\
ALFOSC (NOT)  & H$\alpha$   & 4/5 December 2016 & Imaging \\
\hline
\end{tabular}
\\$^a$: Noise per channel with a velocity width of 0.5 km s$^{-1}$.
\label{smaob}
\end{table*}

\subsection{$L'$-band imaging}\label{naco_obs}

Adaptive optics-assisted imaging in the $L'$ (3.8~$\mu$m) band were obtained in Service Mode using NACO \citep{Lenzen03,Rousset03}, the adaptive optics-assisted near-infrared imager and spectrograph at the Very Large Telescope (VLT), on the night of 18/19 July 2012. The brightness of HH~250-IRS in the infrared allowed us to use the source itself for wavefront sensing, using a dichroic that directed the light in the $IJHK$ bands to the wavefront sensor and transmitted light in the $L'$ band to the camera. The optics used produced an image scale of $0\farcs 027$~per pixel on the $1024 \times 1024$ pixels detector, yielding a field of view of $27 '' \times 27 ''$.

A total of 180 images of 0.175~s exposure time each were obtained and stacked at each point of a dither pattern consisting of 15 positions with a random distribution of telescope pointings within a box $8''$ in size. Flat fields obtained as part of the observatory calibration plan were used in the data reduction. Sky frames were obtained by median combining the dithered images of the field without correcting for telescope offsets. The sky-subtracted, flat-fielded images were then combined into a single frame by coadding the individual images obtained at each telescope pointing, after correcting for the offsets from each position to the next. The final coadded frame thus has an equivalent exposure time of 473~s.

\subsection{$H$- and $K$-band integral field spectroscopy}\label{sinfoni_obs}

Integral field spectroscopy of HH~250-IRS in the $1.5 - 2.4$~$\mu$m range was obtained using SINFONI \citep{Eisenhauer03,Bonnet04}, the adaptive optics-assisted near-infrared integral field spectrograph at the VLT, on the night of 24/25 August 2015. HH~250-IRS is sufficiently bright to be used for tip-tilt correction with SINFONI, and higher order turbulence correction was achieved with the use of the laser guide star facility available at the fourth unit of the VLT where SINFONI is mounted. The image quality thus obtained allowed us to clearly split both components of the binary system, and to extract uncontaminated spectra of each of them. Optics delivering a field of view of $3'' \times 3''$ to the focal plane, and a grism covering the entire $H$ and $K$ bands at a spectral resolution $R \simeq 1500$ were used. Twenty-four exposures of 30~s each were taken by placing the target successively in each of the four quadrants of the integral field unit (IFU) field of view, thus giving a total integration time of 12 minutes. Since the separation between the components of
HH~250-IRS is much smaller than the size of the quadrants this allowed us to confine the target to only one of the quadrants in each exposure, thus removing the need to obtain separate sky exposures. Given that each of the sections of the image slicer of SINFONI partly covers the field of view of two quadrants, the sky background of each exposure was estimated by averaging the exposures obtained two steps before and two steps after it so that there would be no
superposition between the (positive) spectral trace of the object and the (negative) spectral traces of the same object in the subtracted sky frame.

The spectral traces at each slice of the field were determined using dedicated calibration frames that record the trace of a continuum point source for each of the slices. In this way we obtained for each slice the function giving the position of the source along the spatial direction at each wavelength, and defined pseudo-apertures centered on each pixel of the slice along the spatial direction. A wavelength calibration frame was constructed by median-averaging and
sigma-clipping all the science frames to remove the spectral traces of the target from the resulting frame, thus leaving the sky lines only. Sky spectra were extracted for each of the pseudo-apertures defined as described above and individually calibrated in wavelength using the OH airglow emission lines as reference \citep{Oliva92}. Spectra of the sky-subtracted science frames were likewise extracted and wavelength calibrated, and the same procedure was followed for an observation of Hip~97787, a B7V star used for telluric feature removal. The extracted spectra of the telluric reference star were combined into a single 1-D spectrum, and all the extracted spectra of the science frame were divided by it and multiplied by a blackbody spectrum of 13,000~K temperature to provide an approximate relative flux calibration. All the sky-subtracted, wavelength-calibrated, telluric feature-corrected spectra extracted from the science frames
were then combined into a single data cube, from which spectra of both components of the binary at arbitrary apertures and images at arbitrarily selected bands could be extracted in a straightforward manner.

Similar observations were carried out for a binary star previously found in the $L'$-band images and suspected to be physically related to HH~250~IRS (see Sect.~\ref{hh250irs}), which was observed on the nights of 1/2 and 5/6 June 2015. The $0\farcs 14$ separation between both components measured in the $L'$-band image, coupled with the absence of a natural guide star bright enough for tip-tilt correction sufficiently close to the position of the target, prevented us from splitting them even with laser-assisted observations in the integral field datacube. Those observations were therefore carried out without using adaptive optics, and the extracted spectrum of this object is thus the blend of both components. The optics used yielded a larger field of view $8''$ across, with a scale of $0\farcs 25$ per pixel. The observation strategy was the same as for HH~250-IRS, now taking 32 observations of 60~s each on both observing nights, thus resulting in 64 minutes of exposure time.

\subsection{$H$-,$K$-, and $L$-band  spectroscopy}\label{irtf_obs}

Observations were carried out with the 3.0~m NASA Infrared Telescope Facility (IRTF) on Maunakea, Hawaii on the night of 17/18 June 2015, using the recently upgraded SpeX \citep{Rayner03} instrument in the short cross-dispersed (SXD) mode and the long cross-dispersed (short grating setting) mode (LXD), which together cover 0.7~$\mu$m to 4.2~$\mu$m.  The total exposure time in SXD was 24 minutes for HH~250~IRS and the same for the nearby source 2MASS 19212367+1054108.  The total exposure time in LXDs was 12.8 minutes for each of these two sources.  An A0 telluric standard star was observed in both modes for telluric correction, within 0.1 air masses
of the target. An argon lamp was observed for wavelength calibration and a quartz lamp for flat fielding. An arc/flat calibration set was
observed for each target/standard pair. The SpeX data were flat fielded, extracted, and wavelength calibrated using \emph{Spextool} \citep{Cushing04}.

\subsection{$N$- and $Q$- band imaging}\label{visir_obs}

$N$- and $Q$-band imaging of HH~250-IRS was obtained with VISIR \citep{Lagage04}, the thermal infrared camera and spectrograph at the VLT, on the nights of 20/21 July 2015 ($N$-band) and on that same night, and also on 6/7 August and 3/4 September 2015 ($Q$ band). The actual filters used were the B10.7, centered on $\lambda = 10.65$~$\mu$m with a full width at half-maximum of $1.37$~$\mu$m; and the Q2 filter centered on $\lambda = 18.72$~$\mu$m with a full width at half-maximum of $0.88$~$\mu$m. The images were obtained using the chopping and nodding technique to remove the effect of the rapidly changing sky emission and of the difference in the telescope thermal signature pattern due to the slight change of position of the secondary when chopping. A chop throw of $4''$ was used, which kept the target within the $38'' \times 38''$ field of view at all times. Only one chop and nod cycle was obtained, with a chopping frequency of 4~Hz. The total exposure time was 60~s in the $N$ band, and 90~s for each of the observations in the $Q$ band. The images produced by the imaging data reduction pipeline could be directly used for our purposes.

\subsection{H$\alpha$ imaging}\label{Halpha}

Images of HH~250 and its environment through H$\alpha$ and [SII] filters, as well as through a broadband $R$ filter, were obtained using the EMMI imager and spectrograph at the ESO New Technology Telescope (NTT) at La Silla Observatory on 18 August 1994 and can be found at the ESO Science Archive. The years elapsed since those images were taken provide an excellent opportunity to investigate changes having taken place in the region in the intervening time. For this purpose we obtained a ten-minute exposure with the ALFOSC imager and spectrograph at the 2.5~m Nordic Optical Telescope at the Roque de los Muchachos Observatory on the night of 4/5 December 2016, also using an H$\alpha$ filter, which provides a depth similar to that obtained with EMMI at the NTT in 1994.

\subsection{Submillimeter imaging}\label{SMA_obs}

SMA observations toward HH 250 IRS were conducted with the extended configuration at 225 GHz on 27 February and 6 March 2015 and at 351 GHz on 31 March 2015.  The pointing center was $\alpha$(J2000) = $19^{\rm h}21^{\rm m}23\fs1$, $\delta$(J2000) = $10^\circ 54' 04''$. In these observations, six antennae were used. 3C279, Titan, and 1751+096 were observed as bandpass, flux, and gain calibrators, respectively. The flux of 1751+096 at 225 GHz was 2.15 Jy on 27 February 2015 and 2.29 Jy on 6 March 2015, and that at 351 GHz was 1.97 Jy on 31 March 2015. The typical systemic temperature of the observations was 120--300 K on 27 February 2015, 110--180 K on 6 March 2015, and 210--640 K on 31 March 2015. CO~(2--1; 230.538 GHz) and 1.3 mm continuum were observed simultaneously in the 225 GHz observations, and CO (3--2; 345.796 GHz) and 0.9 mm continuum in the 351 GHz observation. 512 channels were assigned to a chunk with a bandwidth of 104 MHz for the CO lines. All the data were calibrated using the MIR software package \citep{Scoville93}. The calibrated visibility data were Fourier-transformed with natural weighting and CLEANed with MIRIAD \citep{Sault95} to produce images. The CO images were generated at a velocity resolution of 0.5 km s$^{-1}$.  The angular resolutions and noise levels of our continuum and CO images are listed in Table~\ref{smaob}.

\begin{table}[t]
\caption{Summary of SMA Imaging Parameters}
\begin{tabular}{lcc}
\hline
 & Synthesized Beam & Noise \\
 & & (mJy Beam$^{-1}$) \\
\hline
1.3 mm & 1\farcs 4 $\times$ 0\farcs 9 (106$^\circ$) & 1.3  \\
0.9 mm & 0\farcs 8 $\times$ 0\farcs 5 (100$^\circ$) & 3.9 \\
CO (2--1) & 1\farcs 3 $\times$ 0\farcs 8 (105$^\circ$) & 150$^a$ \\
CO (3--2) & 0\farcs 9 $\times$ 0\farcs 5 (103$^\circ$) & 210$^a$ \\
\hline
\end{tabular}
\\$^a$: Noise per channel with a velocity width of 0.5 km s$^{-1}$.
\label{smaob}
\end{table}

\begin{figure}[ht]
\begin{center}
\hspace{-0.5cm}
\includegraphics [width=8.5cm, angle={0}]{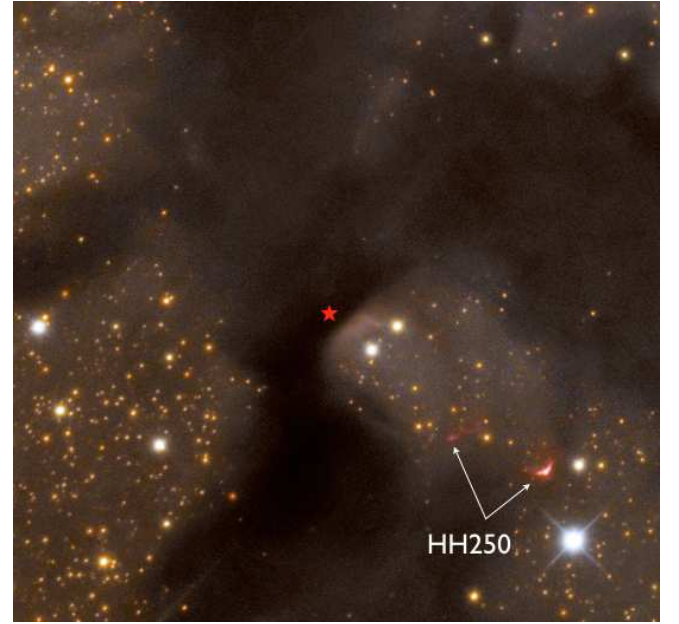}
\caption []{Optical ($BGR$) overview of the star forming region centered on HH~250~IRS = IRAS 19190+1048. The red asterisk marks the embedded source, and two components of the HH~250 flow are marked (A is the brighter knot, B is the fainter one). The image is $5\farcm 7$ across, with north to the top and east to the left. Image courtesy Adam Block.}
\label{panorama}
\end{center}
\end{figure}

\section{Environment and Herbig-Haro flow\label{hh}}

HH~250-IRS, also known as IRAS~19190+1048 (2MASS J19212262+1053569, WISE J192122.62+105357.0) is located in a compact cloud core in the southern part of the large highly structured cloud complex Lynds~673 (Figure~\ref{panorama}; for a review of star formation in the Aquila clouds see \citealt{Prato08}). The galactic coordinates of the source are $l=46.03$, $b=-1.58$, which places it in Cloud~B that is located at higher galactic longitudes than the large Aquila Rift \citep{Dame85}. Cloud~B appears to be a small extension of the Aquila Rift, because of continuity in location and velocity, which places it at a distance of
about 300~pc \citep{Dame87,Sakamoto97}. \citet{Straizys03} determined a distance of $\sim$270~pc to the core of the Aquila Rift, with a distance to the cloud face of $\sim$225~pc. \citet{Prato03} adopted a distance of 150~pc to the nearby T~Tauri star AS~353A. Recently \citet{Ortiz17} found a distance to the neighboring Serpens/W40 complex of 436$\pm$9~pc, suggesting that the Aquila Rift is more distant at lower longitudes. Evidently the distances to objects at
this galactic longitude are poorly known, in the following we adopt the distance of 225~pc by \citet{Straizys03} while recognizing the uncertainty in this value.

\citet{Devine97} discovered a bright compact Herbig-Haro object, labeled HH~250A, near IRAS~19190+1048, and just 14~arcmin southeast of the famous HH~32 \citep{Herbig83}. HH~250A, located at $\alpha_{2000} =$19:21:13.8, $\delta_{2000} = +$10:52:30, has a clear bow shock morphology, facing away from the IRAS source and is located in what appears to be an illuminated outflow cavity stretching away from it (see Figure~\ref{panorama}). A fainter knot (B) is located about $50''$ further to the ENE and near the line joining HH~250A and the IRAS source. At an angular distance of $157''$, IRAS~19190+1048 is the nearest IRAS source to HH~250A and the only embedded source in the cloud. The second nearest source, IRAS~19188+1050, is located $4 \farcm 2$ away, slightly outside the dark cloud in a direction that is almost perpendicular to the axis of the concavity of HH~250A, and the third nearest source is over $9 \farcm 1$ away, outside the cloud and also in a direction far removed from that axis. Given the close proximity to HH~250-IRS, its location close to the direction defined by the axis of the HH~250A bow shock, the morphology of the dark cloud in the area, and the lack of other suitable candidates in the area, we consider that IRAS~19190+1048 is the only realistic candidate driving source of HH~250A and B.

\begin{figure}[ht]
\begin{center}
\hspace{-0.5cm}
\includegraphics [width=8.5cm, angle={0}]{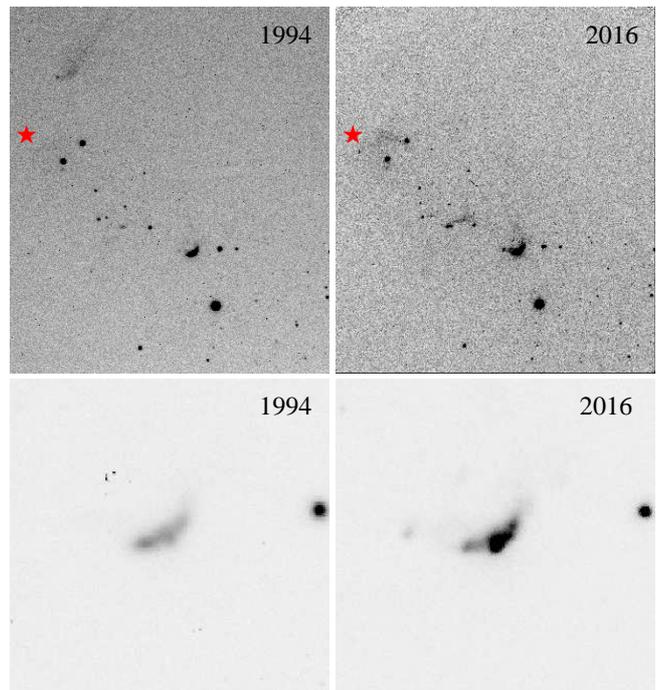}
\caption []{{\sl Upper panels:} The HH 250 region imaged with an H$\alpha$ filter at two different epochs 22 years apart. The 1994 images were obtained with the NTT/EMMI, and the 2016 ones with the NOT/ALFOSC. The field is $4\farcm 0 \times 4\farcm 5$ across. The reflection nebula seen to the NNW of the embedded source has disappeared in the intervening time, whereas nebulosity just E of the source has brightened in the most recent image. The red triangle marks the position of the embedded source HH~250-IRS. {\sl Lower panels:} a close-up view of the bow shock reveals a dramatic brightening of the H$\alpha$ emission between both epochs, as well as other changes in its structure. The field is $41'' \times 41''$ across. The grayscale of the lower panels has been stretched with respect to the upper ones to better show structure in the bow shock.}
\label{HH250_comparison}
\end{center}
\end{figure}

A comparison between the images obtained in 1994 and 2016 is presented in Figure~\ref{HH250_comparison}, where some obvious changes can be easily seen. Figure~\ref{HH250_comparison} shows that the diffuse (presumably reflection) nebulosity NW of HH~250-IRS seen in 1994 has become much fainter or disappeared in 2016, whereas HH~250B has become brighter in the meantime. Another area of nebulosity, close to the embedded IRAS source and near the eastern edge of the image, was barely visible in 1994 but has gained prominence 22 years later. These changes point to a dynamic environment near HH~250-IRS, with dusty clumps in its immediate surroundings causing different parts of the region being alternatively exposed to, or shadowed from, direct illumination by the central source.

Dramatic changes affect HH~250A itself, as clearly seen in the lower panels of Figure~\ref{HH250_comparison}. The arc shape is preserved, but a bright knot has developed near the apsis of the arc in the intervening years and now dominates the appearance of the object. Other subtler changes in the bright emission can be seen, including the appearance of a faint knot downstream along the eastern branch of the arc. The changes in the structure of HH~250A make it difficult to precisely determine its proper motion, but a bulk displacement of the whole structure $\simeq 1\farcs 0$ to the southwest can be measured. This translates into $\simeq 0\farcs 046$/yr, or $\simeq 50$~km~s$^{-1}$ at the adopted distance of the region. The direction of the proper motion away from IRAS~19190+1048 further confirms the identity of the latter as the driving source of HH~250, and the angular distance between both indicates that the Herbig-Haro object is the result of an outburst having taken place about 3,500 years ago.

\section{HH~250-IRS\label{hh250irs}}

HH~250-IRS, the powering source of HH~250, appears resolved into a binary in all our images. The northwestern and southeastern components, hereafter referred to as HH~250-IRS~NW and HH~250-IRS~SE respectively, are separated by a angular distance of $0\farcs 53$ with the SE component located at a position angle of $142^\circ$ with respect to the NW one. At the adopted distance the angular separation corresponds to a projected distance of 120~AU between both components.

\begin{figure}[ht]
\begin{center}
\hspace{-0.5cm}
\includegraphics [width=8.5cm, angle={0}]{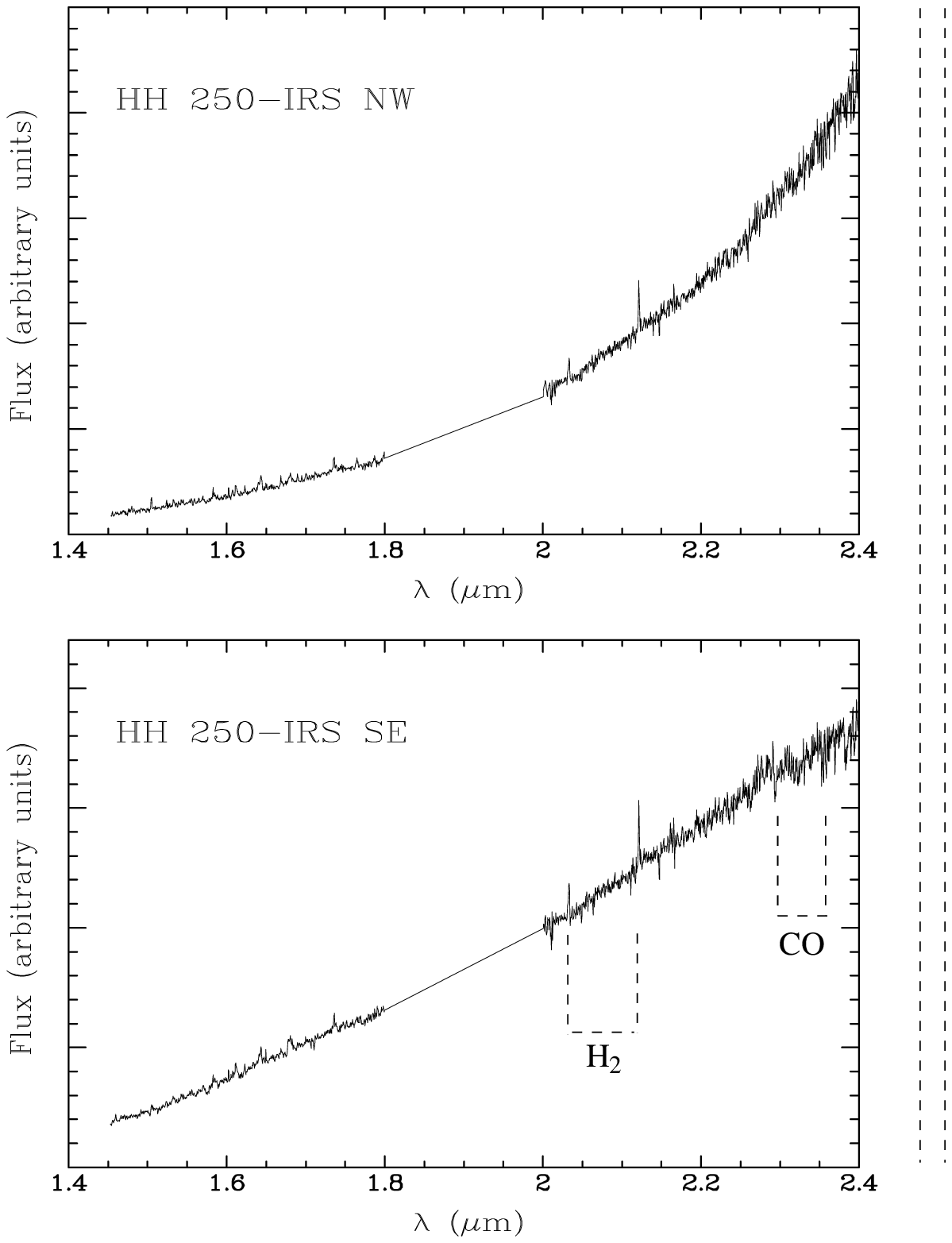}
\caption []{$H-$ and $K-$ band SINFONI spectra of both components of HH~250-IRS. Both sources display a heavily reddened and veiled continuum with emission in the H$_2$ line at 2.122~$\mu$m as the only outstanding feature. HH~250-IRS SE, the least reddened of both components, also shows hints of CO absorption starting at 2.293~$\mu$m.}
\label{spec_SE}
\end{center}
\end{figure}

\subsection{Infrared spectroscopy and photometry\label{infrared}}

The 1.5-2.4~$\mu$m spectra of both components of HH250-IRS are similar and typical of deeply embedded Class~I or flat spectrum sources. They are essentially featureless, with the exception of hints of weak CO absorption longward of 2.293~$\mu$m in the SE component, moderately intense emission lines in both spectra of H$_2$ at 2.122~$\mu$m, and weaker but also clear H$_2$ emission a 2.04~$\mu$m. The non detection of H$_2$ at 2.248~$\mu$m despite the strength of the line at 2.122~$\mu$m indicates that the emission is produced by non dissociative shocks rather than fluorescence. The images obtained by integrating the data-cubes in a narrow spectral range around 2.12~$\mu$m show no traces of extended structure, suggesting that the H$_2$ emission probably arises either from an unresolved region at the base of an outflow, or from an accretion shock. The weak or absent CO absorption is suggestive of a strongly veiled cool photosphere. The SpeX spectrum in the 2.8-4.2~$\mu$m range is dominated by the very strong 3~$\mu$m absorption feature which indicates a substantial amount of water ice along the line of sight \citep[e.g.][]{Graham98}. The 3.45~$\mu$m depression is also clearly visible, it is ascribed to saturated aliphatic hydrocarbons in grains \citep{Pendleton94} and correlates closely with water ice
\citep{Brooke99}. Two emission lines are visible, the Pf$\gamma$ line at 3.74~$\mu$m and the Br$\alpha$ line at 4.05~$\mu$m.

\begin{figure}[ht]
\begin{center}
\hspace{-0.5cm}
\includegraphics [width=8.5cm, angle={0}]{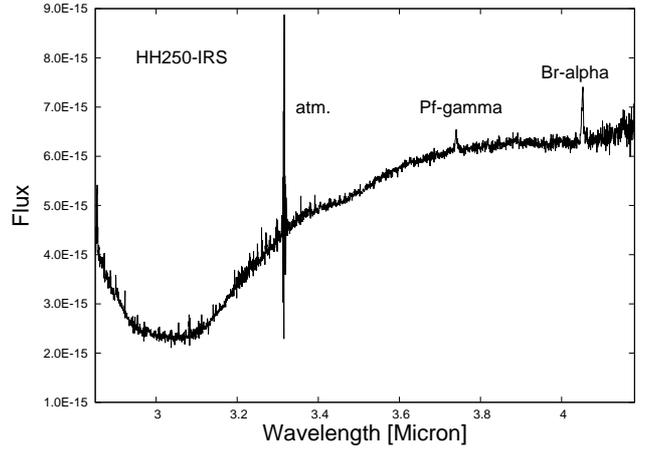}
\caption []{Combined SpeX spectrum of both components of HH~250-IRS
  covering the 2.85~$\mu$m to 4.18~$\mu$m spectral region. The
  Pf$\gamma$ line at 3.7406~$\mu$m and the Br$\alpha$ line at
  4.0523~$\mu$m are prominently in emission. Strong water ice dominates
  the 3~$\mu$m region of the spectrum.}
\label{spec_IRTF}
\end{center}
\end{figure}

\begin{figure}[ht]
\begin{center}
\hspace{-0.5cm}
\includegraphics [width=8.5cm, angle={0}]{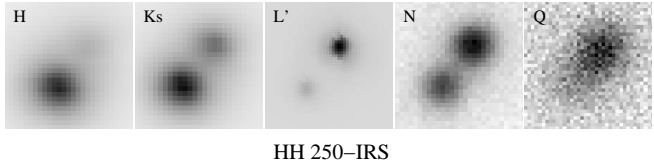}
\caption []{Images of the resolved HH~250-IRS binary from 1.5~$\mu$m to 19~$\mu$m, showing the predominance of the SE components shortward of 3~$\mu$m and of the NE component at longer wavelengths. $H$ and $K_S$ images were obtained with SINFONI, $L'$ with NACO, and $N$ and $Q$ with VISIR. The separation between both components is $0\farcs 53$. North is to the top and east to the left.}
\label{HH250IRS_HKLNQ}
\end{center}
\end{figure}

Figure~\ref{HH250IRS_HKLNQ} shows the binary system in all the bands in which we observed it, between 1.5~$\mu$m and 19~$\mu$m. The figure makes obvious that the NW component has overall redder colors than the SE, but the brightness ratio between both components does not follow a monotonical trend with wavelength. While the SE component dominates in the $H$ and $K$ bands and yields to the NW component at longer wavelengths, the long wavelength dominance of the NW component is greatly reduced in the $N$ band.

We have measured individual magnitudes in the $H$ and $K$ bands by performing photometry on an image obtained by integrating the SINFONI datacube over the wavelength ranges covered by the 2MASS filters in each of those bands. The individual magnitudes of each component were obtained by adding up their fluxes and assuming the sum to be equal to the flux derived from the combined magnitude listed in the 2MASS All-Sky Point Source Catalog \citep{Skrutskie06}. A similar comparison was made to derive the $L'$ magnitudes, now using the instrumental photometry of the NACO image in that band and comparing it to the flux in the $W1$ band listed in the WISE All-Sky Point Source catalog \citep{Wright10}. For the VISIR observation in the $N$ band an observation of the star $\beta$~CrA carried out shortly before the observation of HH~250-IRS was taken as a reference, using the published magnitude of that star ($W3 = 1.348$) as a reference. Finally, for the $Q-$band magnitude we used the WISE catalog values, correcting for the fact that the $Q$ filter used with VISIR has a central wavelength lying between those of the $W3$ and $W4$ bands. The correction was made assuming that the combined flux of both components of HH~250-IRS has a power-law dependence with wavelength, $f \propto (\lambda / \lambda_0)^\alpha$. We thus obtained an interpolated magnitude for the combined components $Q = W3-0.741(W3-W4)$. The noisy VISIR $Q-$band image and the proximity of the diffraction limit of the VLT to the separation between both components of the binary prevented us from carrying out an accurate derivation of the respective contributions, but a rough two-Gaussian fit yielded a magnitude difference $\Delta Q \simeq 1.6$ between both components. Table~\ref{mags} lists the magnitudes thus obtained in each band.

With the exception of the $N-$ band measurement, which was calibrated through the contiguous observation of a star of constant brightness, our derivation of fluxes in all the other bands relies on the cataloged combined fluxes of both components of HH250-IRS, and are thus in principle subjected to uncertainties caused by possible variability. However, the cataloged variability likelihood of HH250-IRS is low in all the WISE bands and consistent with a source of constant brightness, thus ruling out significant errors due to variability in the $L'$, $N$ and $Q$ bands. Since the near-infrared spectra of both components is likely to be dominated by the emission of the circumstellar envelope rather than by photospheric emission, the non variability of the emission at the longer wavelengths suggests that the $H$ and $K_S$ fluxes are also approximately constant. This is also expected on statistical grounds, as the fraction of sources with variability amplitudes exceeding 0.5~mag among disk-bearing young stellar objects in nearby star forming regions does not exceed 3\% \citep{Scholz12}. We thus consider that the bases of the flux calibrations presented here are reliable, and that the features of the spectral energy distributions discussed below are unaffected by the non simultaneity of our observations.

\begin{table}[t]
\caption{Magnitudes of HH~250-IRS}
\begin{tabular}{lcc}
\hline
Band & HH~250-IRS~NW & HH~250-IRS~SE \\
\hline
$H$   & $14.15 \pm 0.1$  & $12.34 \pm 0.05$ \\
$K_S$ & $11.50 \pm 0.05$ & $10.82 \pm 0.05$ \\
$L'$  &  $8.18 \pm 0.05$ &  $9.98 \pm 0.1$  \\
$N$   &  $4.55 \pm 0.1$  &  $4.89 \pm 0.1$  \\
$Q$   &  $1.7  \pm 0.2$  &  $3.3  \pm 0.3$  \\
\hline
\end{tabular}
\label{mags}
\end{table}

\begin{figure}[ht]
\begin{center}
\hspace{-0.5cm}
\includegraphics [width=8.5cm, angle={0}]{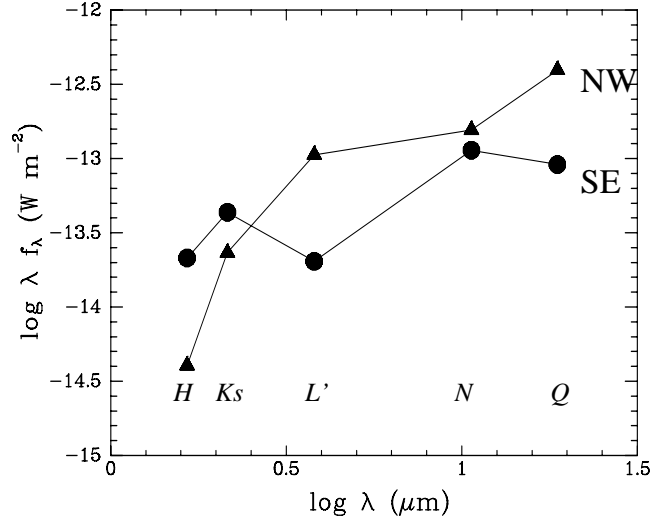}
\caption []{Spectral energy distributions of both components of the HH~250-IRS binary system.}
\label{sed}
\end{center}
\end{figure}

The spectral energy distributions of both components of HH~250-IRS are presented in Figure~\ref{sed}. While both sources show an overall rising shape with wavelength that characterizes them as Class~I sources, in consistency with their spectra, clear differences already inferred from the sequences of images shown in Figure~\ref{HH250IRS_HKLNQ} appear between both. Overall HH~250-IRS~NW is more deeply embedded probably due to the presence of a more massive circumstellar envelope around it. On the other hand, the $N-$band fluxes hint at the presence of strong silicate spectral features in that region appearing in absorption in the NW component and in emission in the SE one. This is consistent with the overall trend observed in the Class~I / Class~II sequence \citep{Kessler05}, where the strong and broad absorption produced by amorphous silicates near 9.7~$\mu$m gradually turns into emission as the circumstellar dust distribution evolves from an envelope to a disk configuration and the spectral energy distribution turns bluer as the central object progressively emerges. It is thus intriguing that the two components of the HH~250-IRS binary system, being essentially coeval, display differences in their spectral energy distributions that are normally associated with the evolutionary stages of their circumstellar environments. The differences between both spectral energy distributions are much above what may be accounted for by the magnitude uncertainies.
Furthermore, whereas HH~250-IRS~NW displays a clear excess flux in the $L'$ band as compared to that expected from the extrapolation of the $H$ and $K_S$ fluxes if produced by a reddened photosphere, the $HK_SL'$ colors of HH~250-IRS~SE are consistent with a reddened photosphere with little infrared excess, which manifests itself only at longer wavelengths. This points to a lack of warm dust in the immediate vicinity of the central source of HH~250-IRS~SE, hinting at the existence of a dust-free cavity around the central star of HH~250-IRS~SE that has no analog in the envelope of the NW component. Large inner holes in the distribution of dust are the hallmark of transition disks, which are usually thought to be caused by circumstellar dust evolving from optically thick to optically thin due to grain growth, formation of planetesimals, and eventual clearance of dust from the inner regions of disks by newly formed planets \citep[e.g.][]{Cieza10}. It would be surprising that such processes had had time to operate in a system as young as HH~250-IRS, and we prefer instead to relate the existence of the gap in the inner disk of HH~250-IRS~SE to the presence of the NW binary component, as we describe in Sect.~\ref{discussion}.

\subsection{Cold material around HH~250-IRS\label{cold}}

\subsubsection{CO Emission}\label{co}

\begin{figure*}[ht]
\begin{center}
\hspace{-0.5cm}
\includegraphics [width=14cm, angle={0}]{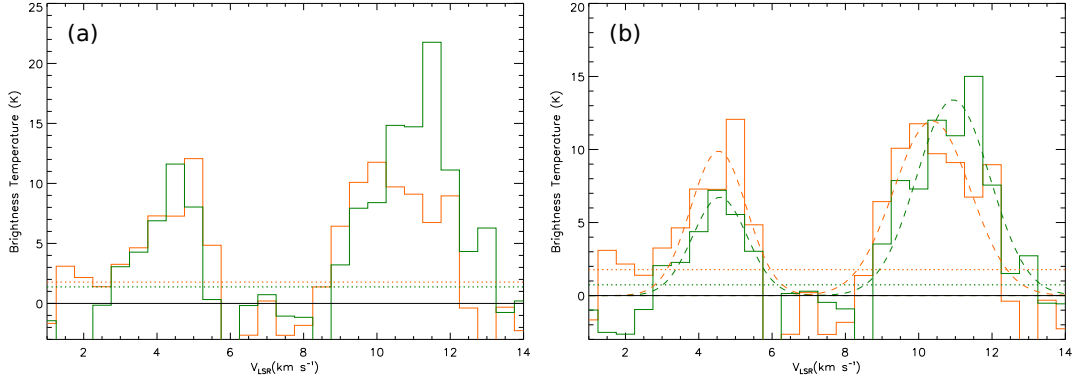}
\caption []{CO spectra obtained with our SMA observations. Orange and green histograms present the CO (2--1) and (3--2) emission, respectively. The spectra in both panel (a) and (b) were extracted by averaging over the region with a radius of 0$\farcs 5$ centered on the continuum peak position from the images with the correction of the primary beam attenuation. Dashed lines denote 1$\sigma$ noise levels of the spectra. In panel (b), the CO (3--2) spectrum was extracted after convolving the CO (3--2) image with the same angular resolution of the CO (2--1) image. Red and blue curves show the Gaussian fitting results of the line profiles.}
\label{cospec}
\end{center}
\end{figure*}

\begin{figure*}[ht]
\begin{center}
\hspace{-0.5cm}
\includegraphics [width=14cm, angle={0}]{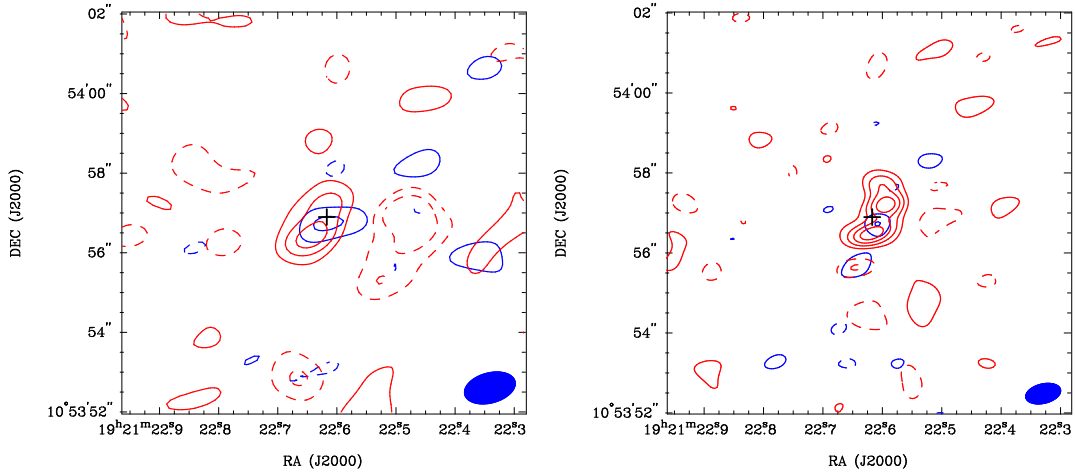}
\caption []{Integrated intensity maps of the CO (2--1; left) and (3--2; right) emission obtained with our SMA observations. Blue and red contours present the blueshifted and redshifted velocity components centered at $V_{\rm LSR} \sim 4.6$ km s$^{-1}$ and $V_{\rm LSR} \sim 10.7$ km s$^{-1}$, respectively. The integrated velocity widths of the blueshifted and redshifted components are 1.5 km s$^{-1}$ and 3.5 km s$^{-1}$ in the CO (2--1) emission, and those are 1 km s$^{-1}$ and 3 km s$^{-1}$ in the CO (3--2) emission. Crosses denote the continuum peak position, and filled blue ellipses show the sizes of the synthesized beams. Contours are from 3$\sigma$ in steps of 2$\sigma$. 1$\sigma$ noise levels in the CO (2--1) maps of the blueshifted and redshifted components are 0.14 and 0.2 Jy Beam$^{-1}$ km s$^{-1}$, and those are 0.15 and 0.24 Jy Beam$^{-1}$ km s$^{-1}$ in the CO (3--2) maps.}
\label{comap}
\end{center}
\end{figure*}

Figure \ref{cospec}a presents the spectra of the CO (2--1) and (3--2) emission lines observed with the SMA. The spectra are extracted at the continuum peak position and averaged over the region within a radius of 0\farcs5.  The area for averaging corresponds to the area of the synthesized beam of the CO (2--1) image. Two velocity components are detected in the CO spectra, one centered at $V_{\rm LSR} \sim 4.6$~km~s$^{-1}$ (hereafter blueshifted component) and one at $V_{\rm LSR} \sim 10.7$~km~s$^{-1}$ (hereafter redshifted component). Figure~\ref{comap} shows the integrated intensity maps of the two velocity components detected in the CO lines. In both the CO (2--1) and (3--2) emission, the redshifted component is extended and elongated along the northwest-southeast direction with a position angle of $\sim$150$^\circ$.  This orientation is perpendicular to the outflow direction and matches well the line between components SE and NW of HH~250~IRS. The blueshifted component is not resolved and has a size comparable to the angular resolution. Because of the limited signal-to-noise ratios, the velocity structures of the CO (2--1) and (3--2) emission of both components are barely revealed.

To compare the brightness temperature $T_b$ of the CO (2--1) and (3--2) emission we convolved the CO (3--2) image with the
angular resolution of the CO (2--1) image after correcting for primary beam attenuation. The spectra of the CO (2--1) and (3--2) emission extracted at the same angular resolutions are presented in Figure~\ref{cospec}b. By fitting a Gaussian line profile to the spectra, we measured their peak brightness temperatures, the centroid velocities in the LSR frame, and the FWHM line widths. The results of the fit are listed in Table \ref{specfit}. For each velocity component, we assume that the CO (2--1) and (3--2) emission lines trace the same bulk of gas, although there is an offset of 0.5 km s$^{-1}$ in the centroid velocities of their redshifted component. We then used the radiative transfer code RADEX \citep{Vandertak07} to simultaneously fit the observed brightness temperatures in both CO transitions with the observed line widths. We performed spectral fits with various fixed values of the volume density $n_{\rm H_2}$ and varied the kinetic temperature $T_k$ and the column density $N_{\rm co}$ to obtain a predicted $T_{\rm b}$ to be compared with the actually observed values, thus obtaining the best-fit excitation temperature $T_{\rm ex}$, optical depth $\tau$, and $T_k$. We considered $n_{\rm H_2} > 10^5$~cm$^{-3}$, typical of the average gas density on a 100~AU scale around young stellar objects. The gas temperature $T_{\rm k}$ is estimated to be 13-18 K for the blueshifted component and 24-40 K for the redshifted component. The CO lines in both components are thermalized. The CO (2-1) and (3-2) lines in the blueshifted component are both optically thick, yielding only a lower limit to the CO column density. In the redshifted component, the CO (2-1) line is optically thin to marginally optically thick, and the CO (3-2) line is close to optically thick. Its CO column density is estimated to be $(2 - 3) \times 10^{16}$ cm$^{-2}$.

%  From Figure \ref{comap}, the area of the extended CO emission detected at more than 3$\sigma$ level corresponding to the redshifted component is measured to
%be 3.3 arcsec$^2$. Assuming a roughly uniform CO column density within this area, the gas mass in this component is estimated to be $\sim 1 \times %10^{-5}$~M$_\odot$ with the typical CO abundance of $N_{\rm CO} / N_{\rm H_2}$ = 10$^{-4}$ \citep{Lacy94} and a mean molecular weight of 2.33. The mass of the %blueshifted component is more difficult to estimate, as it is unresolved; assuming a maximum radius of $\simeq 0\farcs 3$ that would still make it appear %unresolved, a crude upper limit of the mass is obtained as $\sim 2 \times 10^{-6}$~M$_\odot$.

%That is more than two orders of magnitude
%lower than the mass estimated from the continuum emission assuming the gas-to-dust mass ratio of 100.

\begin{table*}[t]
\caption{Gaussian fitting to the CO (2--1) and (3--2) spectra}
\begin{tabular}{lccccccc}
\hline
 & \multicolumn{3}{c}{Blueshifted Component} && \multicolumn{3}{c}{Redshifted Component} \\
\cline{2-4} \cline{6-8}
\hline
Line & $T_{\rm b}$ & $V_{\rm c}$ & $\Delta V$ && $T_{\rm b}$ & $V_{\rm c}$ & $\Delta V$ \\
 & (K) & (km s$^{-1}$) & (km s$^{-1}$) && (K) & (km s$^{-1}$) & (km s$^{-1}$) \\
CO (2--1) & 9.9$\pm$1.2 & 4.5$\pm$0.1 & 1.8$\pm$0.3 && 11.8$\pm$1.1 & 10.4$\pm$0.1 & 2.5$\pm$0.3 \\
CO (3--2) & 6.7$\pm$0.6 & 4.6$\pm$0.1 & 1.8$\pm$0.2 && 13.4$\pm$0.5 & 10.9$\pm$0.1 & 2.4$\pm$0.1 \\
\hline
\end{tabular}
\label{specfit}
\end{table*}

\begin{table*}[t]
\caption{Gas temperature estimated from the CO (2--1) and (3--2) spectra}
\begin{tabular}{cccccccccc}
\hline
  & & & \multicolumn{3}{c}{CO (2--1)} && \multicolumn{3}{c}{CO (3--2)} \\
\cline{4-6} \cline{8-10}
$n_{\rm H_2}$ & $T_{\rm k}$ & $N_{\rm co}$ & $T_{\rm ex}$ & $\tau$ & $T_{\rm b}$ && $T_{\rm ex}$ & $\tau$ & $T_{\rm b}$ \\
(cm$^{-3}$) & (K) & (cm$^{-2}$) & (K) & & (K) && (K) & & (K) \\
\multicolumn{10}{c}{Blueshifted Component} \\
\hline
10$^{5}$ & 15$^{+4}_{-1}$ & $>$1.2 $\times$ 10$^{16}$ & 15$^{+3}_{-1}$ & $>$0.9 & 9.0$^{+0}_{-1.0}$ && 14$^{+4}_{-1}$ & $>$1.1 & 6.9$^{+0}_{-0.7}$ \smallskip\\
$\geq$10$^{6}$ & 14$^{+4}_{-1}$ & $>$1.3 $\times$ 10$^{16}$ & 14$^{+4}_{-1}$ & $>$1 & 8.8$^{+0}_{-0.8}$ && 14$^{+4}_{-1}$ & $>$1.1 & 7$^{+0.2}_{-0.7}$ \smallskip\\
\hline
\multicolumn{10}{c}{Redshifted Component} \\
\hline
10$^{5}$ & 32$^{+8}_{-6}$ & (2.3$^{+0.8}_{-0.3}$) $\times$ 10$^{16}$ & 31$^{+8}_{-6}$ & 0.6$^{+0.5}_{-0.2}$ & 11.7$^{+1.6}_{-1.3}$ && 30$\pm$6 & 0.9$^{+0.6}_{-0.3}$ & 13.4$^{+0.3}_{-0.1}$ \smallskip \\
$\geq$10$^{6}$ & 28$^{+8}_{-4}$ & (2.5$^{+0.8}_{-0.4}$) $\times$ 10$^{16}$ & 28$^{+8}_{-4}$ & 0.7$^{+0.5}_{-0.3}$ & 11.8$^{+1.4}_{-1.5}$ && 28$^{+8}_{-4}$ & 1.1$^{+0.2}_{-0.4}$ & 13.3$\pm$0.1 \smallskip \\
\hline
\end{tabular}
\\$n_{\rm H_2}$ is the H$_2$ number density, $T_{\rm k}$ is the gas kinematic temperature, $N_{\rm co}$ is the CO column density, $T_{\rm ex}$ is the excitation temperature, $\tau$ is the optical depth, and $T_{\rm b}$ is the peak brightness temperature. In our spectral fitting with the radiative calculation, $T_{\rm k}$ and $N_{\rm co}$ are the free parameters, and $n_{\rm H_2}$ is a fixed parameter.\label{coradex}
\end{table*}

\subsubsection{Continuum Emission \label{continuum}}

\begin{figure*}[ht]
\begin{center}
\hspace{-0.5cm}
\includegraphics [width=14cm, angle={0}]{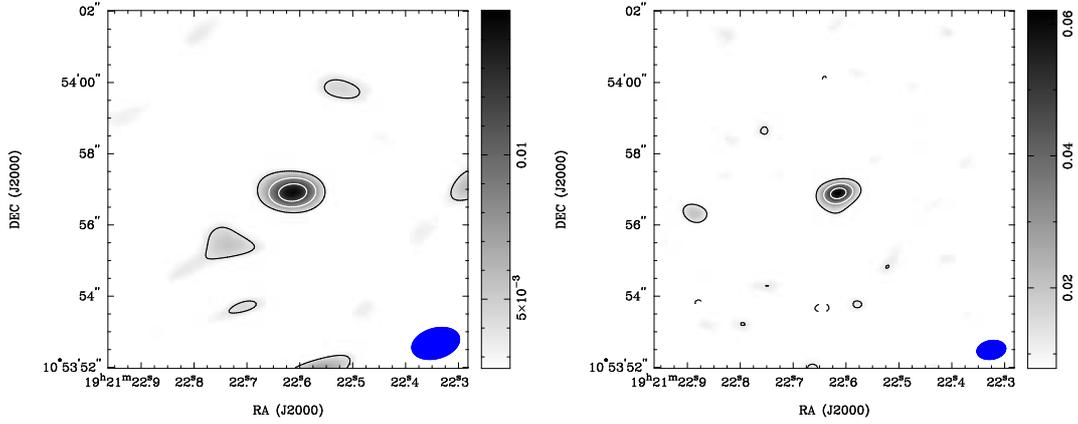}
\caption []{1.3 mm (left) and 0.9 (right) continuum images of HH 250-IRS A obtained with our SMA observations. The images are not corrected for the primary beam attenuation. Contour levels start from 3$\sigma$ in steps of 3$\sigma$ in the 1.3 mm image and in steps of 5$\sigma$ in the 0.9 mm image, where 1$\sigma$ is 1.3 and 3.9 mJy Beam$^{-1}$, respectively. Filled blue ellipses show the sizes of the synthesized beams.}
\label{conti}
\end{center}
\end{figure*}

Figure \ref{conti} presents the 1.3 mm and 0.9 mm continuum images of HH 250-IRS. The peak position is measured to be $\alpha(2000) = $19:21:22.62, $\delta(2000) = $10:53:56.9 at both wavelengths, which is within about 0$\farcs$1 of the 2MASS position. The apparent full-width-half-maximum (FWHM) sizes of the continuum emission are measured to be 1\farcs 4 $\times$ 0\farcs 9 at 1.3 mm and 0\farcs 8$\times$ 0\farcs 6 at 0.9 mm by fitting a two-dimensional Gaussian distribution with the MIRIAD task {\it imfit}. The observed sizes are consistent with the angular resolutions, and the deconvolution by {\it imfit} results in a point source. By fitting a point source to the visibility data, the flux is measured to be 17$\pm$1.4 mJy at 1.3 mm and 81.5$\pm$5.6 mJy at 0.9 mm. The spectral index $\alpha$ of the continuum emission between 1.3 mm and 0.9 mm is derived to be 3.5$\pm$0.2, and the frequency dependence of the dust opacity $\beta = \alpha - 2$ is 1.5$\pm$0.2.

The mass of the circumstellar dust traced by the continuum emission ($M_{\rm dust}$) can be estimated as

$$M_{\rm dust} = \frac{F_{\lambda}d^{2}} {\kappa_{\lambda} B(T_{\rm dust})} \eqno(1),$$

\noindent where $F_{\lambda}$ is the continuum flux at the wavelength $\lambda$, $d$ is the distance to the source, $\kappa_\lambda$ is the dust mass opacity at the wavelength $\lambda$, $T_{\rm dust}$ is the dust temperature, and $B(T_{\rm dust})$ is the Planck function at a temperature of $T_{\rm dust}$. By adopting the frequency function for dust mass opacity, $\kappa_\lambda = 10 \times (0.3\ {\rm mm}/\lambda)^\beta$ cm$^2$ g$^{-1}$ \citep{Beckwith90}, we obtain
$\kappa_{\rm 1.3mm} = (1.1 \pm 0.3)$~cm$^2$~g$^{-1}$ and $\kappa_{\rm 0.9mm} = (1.9 \pm 0.6)$~cm$^2$~g$^{-1}$ with the measured $\beta$ value\footnote{It should be noted that we use the opacity per unit dust mass, rather than the more commonly used opacity per unit total mass, since we only estimate the dust mass here.}

Given the coincidence of the peak of the continuum emission with the peak of the blueshifted component of the CO emission, both being unresolved, it appears reasonable to assume that both trace the same structure. Therefore we adopt the derived temperature of the dust as 14~K, assuming that dust and gas are thermally well coupled. In this way, $M_{\rm dust}$ is estimated to be $(1.6 \pm 1.1) \times 10^{-4}$~M$_\odot$ from the 1.3~mm measurement, and $(2.1 \pm 0.9) \times 10^{-4}$~M$_\odot$ from the 0.9~mm measurement. Assuming a standard gas-to-dust ratio of 100 by mass \citep{Draine07} this points to a rather massive envelope in excess of $10^{-2}$~M$_\odot$. However, the crude assumption of a single dust temperature, existing evidences for gas depletion in circumstellar disks \citep{Ansdell16}, and the lack of an independent estimate of the gas mass of the envelope through an optically thin tracer refrains us from making any definitive claims in this regard.

\section{Discussion\label{discussion}}

\subsection{A circumbinary disk around HH~250-IRS \label{disc-circumbinary}}

The distinct kinematical components detected in CO emission, together with their different spatial distributions and temperatures, suggest that they trace different structures in the HH~250-IRS system. The CO emission is clearly elongated in the direction defined by both components of the binary, and shows hints of a double-peaked spatial distribution of the intensity in the CO (3--2) map with both peaks separated by an angular distance of $\sim 1\farcs 1$ that exceeds the angular distance between the components of the binary. This leads us to assume that the CO emission associated with the redshifted component probably comes from an extended, torus-shaped circumbinary envelope surrounding both members of the HH~250-IRS system, with a dimension along the diameter of the torus of $\sim 250$~AU. The unresolved blueshifted gas component that we attributed in Sect.~\ref{continuum} to a circumstellar envelope around one of the components of HH~250-IRS seems most likely related to HH~250-IRS~NW, the most embedded member of the system displaying the strongest excess at mid-infrared wavelengths. However, the close coincidence between the peak position of the continuum emission with the 2MASS position rather points to HH~250-IRS~SE, the brightest member of the pair at the 2MASS bands, as the dominant source at millimeter wavelengths. Higher-resolution observations with accurate astrometry are needed to clarify the association.

A crude estimate of the mass of the circumbinary disk can be obtained from the CO column density derived from the
spectral fit (Table~\ref{coradex}), integrating it over the angular area of the redshifted component ($3.3$~arcsec$^2$), and assuming a H$_2$-to-CO ratio $N_{\rm H_2} / N_{\rm CO} \simeq 10^4$ \citep{France14} and a mean molecular weight of $2.33 m_H$. In this way we obtain $M \sim 1.7 \times 10^{-5}$~M$_\odot$. The non detection of continuum emission associated with the circumbinary ring sets in turn an upper limit of $M_{\rm dust} \lesssim 10^{-5}$~M$_\odot$. Given the estimated gas mass, the non detection of the circumbinary ring in the continuum is thus expected for a normal gas-to-dust ratio of 100.

Binary systems where both circumstellar and circumbinary disks coexist are a predicted outcome of the fragmentation process leading to multiple star formation when the accreting material possesses a high angular momentum \citep{Bate97}. The circumstellar disks around each component are truncated due to dynamical interaction with the companion, and their long lifetimes inferred from observed systems requires their sustained feeding through accretion streams from the circumbinary disk that acts as a reservoir \citep{Mathieu00}. The material accreted by each member of the binary greatly depends on the mass ratio of the protostellar seeds and the specific angular momentum of both the accreting gas and the secondary. This can lead to widely different accretion paths for both components of the binary, making possible that only one of the components is deeply embedded in a circumstellar envelope. The remarkably different spectral energy distributions of HH~250-IRS NW and SE, and the detection of only one of the circumstellar envelopes in our observations, probably reflects this.

Simulations of the dynamical interaction between a binary system and a massive circumbinary disk show that the latter strongly influences the orbital characteristics of the former \citep{Artymowicz91}, with the production of a highly eccentric system as one of the main outcomes. This leads to the expectation that HH~250-IRS NW and SE may undergo periodic episodes of proximity much closer than the present 120~AU of projected distance, causing major disturbances to their respective disks and possibly major events of mass loss. It is possible that a past periastron passage may have been responsible for the formation of HH~250. Assuming a combined mass of ~$\sim 1$~M$_\odot$ for the binary and a semimajor orbital axis of half the current projected separation of 120~AU, we obtain an orbital period of $\sim 500$ years, which within the significant uncertainties compares quite favorably with the kinematic age of 3,500~yr of HH~250 (Sect.~\ref{hh}) that sets an upper limit to the time since the last outburst. We speculate that major perturbations in the circumstellar disk structure might also be responsible for the cavity around HH~250-IRS hinted by its drop in flux at $3.8$~$\mu$m. Evidence of companions causing large dust-free cavities that masquerade as transition disks has been reported by \citet{Ruiz16}, showing that this mechanism is at work in other binary systems.

\subsection{A comparison with other similar systems \label{disc-comparison}}

The observations presented here indicate that HH~250-IRS is an addition to the handful of binary or multiple systems known to harbor both circumstellar and
circumbinary disks. The best studied of such systems is GG~Tau, where both circumstellar disks and the circumbinary disk have been observed in much detail; see \citet{Dutrey16} for a recent review. That system is probably older than HH~250-IRS, with an age of 1.5~Myr derived from the fit of pre-main sequence evolutionary tracks to the properties of GG~Tau~Aa and Ab. Similarly to HH~250-IRS only one of the two components, GG~Tau Aa, displays submillimeter continuum emission. The inner radius of the circumbinary disk is 180~AU, comparable to the size of the HH~250-IRS circumbinary disk when taking into account the distance uncertainty of the latter.

UY~Aur \citep{Close98, Tang14} is another binary, or possibly triple system, with a projected separation of 120~AU between its two main components \citep{Hartigan03} and a circumbinary disk with a radius of 520~AU. High resolution submillimeter images presented by \citet{Tang14} show that its circumbinary disk is well detected in CO but only marginally in dust continuum. Like in GG~Tau, the photospheres of the UY~Aur binary are well observed, suggesting an older age than that of HH~250-IRS.

A closer and perhaps even younger analog to HH~250-IRS that has attracted recent attention is the L1551~NE system \citep{Reipurth02,Takakuwa12}, a system containing two Class~I members with a projected separation of 70~AU, one of which drives an outflow traced by [FeII] emission, and surrounded by a circumbinary disk with a radius of 300~AU \citep{Takakuwa14}. High resolution images of L1551~NE resolving both the circumbinary and circumstellar disks obtained with ALMA, and a comparison with numerical simulations of the gas flow in the system, are presented by \citet{Takakuwa17}.

Other systems confirmed or suspected to have circumbinary disks are listed in \citet{Monin07} and \citet{Dutrey16}. The best studied example where a well developed jet exists is HH~30, a system where the binarity of the driving source and its properties (masses and separation) are inferred from the wiggling of the jet, rather than by direct observation due to its blocking by a disk seen nearly edge-on \citep{Estalella12}. This makes HH~250-IRS a rare example among the selected group of binary systems surrounded by circumbinary disks, where the connection of the observed characteristics of the system with its outflow activity in the recent past may be more directly established.

\section{Conclusions}

In this paper we report the first dedicated study of HH~250-IRS, the driving source of HH~250, a little studied Herbig-Haro object in Aquila. Our conclusions are summarized as follows:

\begin{itemize}

\item Obvious evolution is observed in the region around HH~250-IRS and HH~250 by comparing H$\alpha$ images obtained in 1994 and 2016. The appearance of the most prominent feature, the arc-shaped HH~250A, is dominated by a bright spot in the 2016 image that was inconspicuous in 1994. The overall proper motion of HH~250A is estimated to be 0$\farcs 046$ yr$^{-1}$ (corresponding to $\sim 50$~km~s$^{-1}$) pointing away from HH~250-IRS and confirming its origin in that source. Different parts of the reflection nebulosity in the surroundings of HH~250-IRS have faded or brightened over that time span, suggesting a clumpy envelope around the illuminating source.

\item The powering source of HH~250-IRS is shown to be a binary system separated by $0\farcs 53$. Both components have Class~I spectral energy distributions, but clear differences appear between both. The near-infrared spectrum of the SE component is closer to the Class I/II transition, and the spectral energy distribution suggests a strong silicate feature in emission, while it seems to be in absorption in the NW component. A drop in the $L'$-band flux of the SE component is tentatively attributed to the dynamical interaction with the NW component.

%\item A close double star ($0\farcs 14$ separation) projected 21'' to the Northeast of HH~250-IRS is shown to have a combined spectrum typical
%  of a cool giant or supergiant star, thus indicating that it is a background source unrelated to the HH~250 region.

\item Millimeter-wave observations reveal two kinematical components of the gas traced by CO emission separated by a difference of $\simeq 6$~km~s$^{-1}$ in radial velocity. The redshifted component is resolved, elongated along the same direction as that defined by the HH~250-IRS NW-SE pair, shows hints of being doubly-peaked, and has a temperature of $\simeq 30$~K. The blueshifted component remains unresolved by our observations and has a temperature of $\simeq 14$~K. The extent and hints of structure of the redshifted component suggest that it traces a circumbinary disk around the HH~250-IRS system, while the blueshifted component may correspond to a circumstellar disk around one of the two components of the binary system.

\item Cold dust traced by continuum emission is detected as an unresolved source centered on HH~250-IRS, being spatially coincidental with the blueshifted component seen in CO emission.

\end{itemize}

The intriguing characteristics of the HH~250-IRS binary system described in this work suggest its potential for the study of accretion from a circumbinary disk onto circumstellar disks in a system in which both components have in turn a strong dynamical interaction. At the same time, the existence of a large-scale outflow revealed by the existence of the HH~250 bow shock offers the possibility of further investigating the link between binarity and collimated outflows. While the observations presented here are in many respects tantalizing due to their limited resolution and sensitivity, higher quality observations can make HH~250-IRS a cornerstone for the understanding of the interplay between the central binary, its circumstellar envelopes, the circumbinary material and the jets. Given the small number of such systems known to date and the diversity of outcomes resulting from the possible combinations of initial parameters that simulations predict, HH~250-IRS is a precious addition to the sample.

\begin{acknowledgements}

We thank the staff at ESO's Cerro Paranal Observatory for their competent execution of our VLT observations in
Service Mode, as well as to the astronomers at ESO's User Support Department for their careful review and expert advice at the time of defining the observations with NACO, SINFONI and VISIR. We are thankful to Anlaug Amanda Djupvik at the NOT for making the arrangements that allowed us to obtain the new H$\alpha$ image presented in this paper, to Grigori Fedorets for carrying out the NOT observations, and to Adam Block for permission to use the image shown in Figure~1
(http://adamblockphotos.com). We also thank the anonymous referee for constructive suggestions. FC wishes to acknowledge the hospitality of the Center for Astrobiology (INTA-CSIC) at Villafranca del Castillo (Spain), especially of Nuria Hu\'elamo, and the generous support granted by the Faculty of the European Space Astronomy Centre (ESAC) for a research stay during which part of this work was carried out.  The Submillimeter Array is a joint project between the Smithsonian Astrophysical Observatory and the Academia Sinica Institute of Astronomy and Astrophysics and is funded by the Smithsonian Institution and the Academia Sinica. We acknowledge the support of the NASA Infrared Telescope Facility, which is operated by the University of Hawaii under contract NNH14CK55B with the National Aeronautics and Space Administration. This work is based in part on observations made with 2MASS and WISE. The Two Micron All Sky Survey (2MASS) is a joint project of the University of Massachusetts and the Infrared Processing and Analysis Center/California Institute of Technology, funded by the National Aeronautics and Space Administration and the National Science Foundation. The Wide-field Infrared Survey Explorer is a joint project of the University of California, Los Angeles, and the Jet Propulsion Laboratory/California Institute of Technology, funded by the National Aeronautics and Space Administration. This research has made use of the SIMBAD database, operated at CDS, Strasbourg, France, and of NASA's Astrophysics Data System Bibliographic Services, as well as of the ESO Science Archive.

\end{acknowledgements}

\bibliographystyle{aa} % style aa.bst
\bibliography{hh250_cit}

\begin{appendix}

\section{An Adjacent Binary\label{false_companion}}

\begin{figure}[ht]
\begin{center}
\hspace{-0.5cm}
\includegraphics [width=8.5cm, angle={0}]{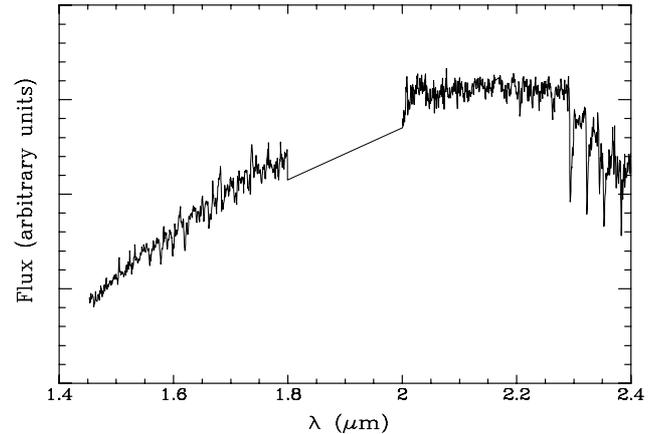}
\caption []{SINFONI spectrum of the close double star located 21$''$
  northeast of the HH~250-IRS binary, initially suspected to be
  physically related to the system. Despite the low photospheric
  temperature indicated by the appearance of the NaI (2.206~$\mu$m)
  and CaI (2.262~$\mu$m) lines and the deep CO bands, the spectrum is
  distinctly different from that of a low-mass young stellar object,
  being instead typical of a late-type giant or supergiant star.
  The apparent emission at 2.166~$\mu$m is an artifact due to the
  presence of Br$\gamma$ absorption in the spectrum of the star used
  for telluric features correction.}
\label{otherbinary}
\end{center}
\end{figure}

While examining our $L'$-band images of HH~250-IRS we noticed another binary 21$''$ to the northeast, which is identified as 2MASS 19212367+1054108 (hereafter 2M192123). This is a fainter object (K=10.56) than HH~250-IRS, and a tighter binary with a separation of only 0$\farcs 15$. Given the proximity to HH~250-IRS in a very high-extinction region, we decided to examine the possibility that 2M192123 and HH~250-IRS might form a bound $\epsilon$-Lyrae type quadruple system in the early stages of formation. We therefore obtained the same kind of spectral data of 2M192123 as for HH~250-IRS. Figure~\ref{otherbinary} shows a SINFONI spectrum of the combined light from the two components of 2M192123. The spectrum clearly indicates a cool photosphere, with recognizable absorption features due to NaI (2.206~$\mu$m), CaI (2.262~$\mu$m) and, most notably, CO. However, the spectrum rules out the possibility that 2M192123 may be a low-mass young stellar object, as the broad water steam absorption wings that should dominate the $H$- and $K$-band spectrum in that case are clearly absent. Furthermore the depth of the CO absorption, for which we measure an equivalent width of 27~\AA\ in the $v$(2--0) band starting at 2.293~$\mu$m, indicates a very low surface gravity characteristic of a supergiant \citep{Lancon00}, over two orders of magnitude lower than for a young stellar object. Finally, WISE observations do not show any evidence for infrared excess in the 3.5-20~$\mu$m range\footnote{The AllWISE Data Release lists W3 and W4 magnitudes which, taken at face value, would indicate substantial infrared excess, $(W2-W3) = 1.55$ and $(W3-W4)=5.62$. However, inspection of the WISE images shows that the object is undetected by WISE in both the W3 and W4 bands, and that the listed values are actually contaminated, most probably by HH~250-IRS itself.}. This is
further strengthened by the complete absence of any detection in the 225 and 351 GHz SMA map of the region. We conclude that 2M192123 is a background object, probably a binary containing at least one M supergiant member, unrelated to the HH~250-IRS binary.

\end{appendix}

%---------------------------------------------------------
% Edit the file globule_ref.bib to update the references.
% Then run the commands:
%       latex globule
%       bibtex globule
%----------------------------------------------------------

%\bibliography{globule_ref}

%\Online

%\listofobjects

\end{document}